\documentclass[pra,superscriptaddress,twocolumn,nofootinbib]{revtex4-1}
\usepackage{graphicx,amssymb,amsmath,color}

\usepackage[colorlinks]{hyperref}

\begin{document}

\title{Tan's contact in a cigar-shaped dilute Bose gas}

\author{Jean Decamp}
\author{Mathias Albert}
\author{Patrizia Vignolo}
\affiliation{Universit\'e C\^ote d'Azur, CNRS, Institut de Physique de Nice, France}

\date{\today}

\begin{abstract}
  We compute the Tan's contact of a weakly interacting Bose gas at zero temperature in a cigar-shaped configuration. Using an effective one-dimensional 
  Gross-Pitaevskii equation and Bogoliubov theory, we derive an analytical formula that interpolates between the three-dimensional and the one-dimensional 
  mean-field regimes. In the 
  strictly one-dimensional limit, we compare our results with Lieb-Liniger theory. Our study can be a guide for actual experiments interested in the study of Tan's 
  contact in the dimensional crossover.
\end{abstract}

\maketitle

%%%%%%%%%%%%%%%%%%%%%%%%%%%%%%%%%%%%%%%%%%%%%%%%%%%%%%%%%%%%%%%%%%%%%%

\section{Introduction}
\label{sec:Intro}
Ultracold atomic gases are a perfect playground to study quantum systems from the non-interacting regime to the strongly interacting one, in one, two, and three dimensions. Both interactions and dimensionality drive correlations in the systems, and, thus, play a crucial role in the off-diagonal 
elements of the one-body density matrix or, equivalently, in its Fourier transform, namely the momentum distribution.
In particular, for $\delta$-interacting systems, Tan's contact $\mathcal{C}$, which determines the asymptotic behavior of the momentum distribution 
$\mathcal{C}=\lim_{k\rightarrow\infty}n(k)k^4$, embeds information on the interaction energy, the density-density correlation function \cite{Tan2008a,
Tan2008b,Tan2008c}, and for fermionic and/or bosonic mixtures it is univocally related to the symmetry, and thus the magnetization, of the system 
\cite{Decamp2016b,Decamp2017}. Moreover, in the case of a three-dimensional (3D) dilute Bose gas, the measurement of the contact has been exploited to access the quantum 
depletion \cite{Clement2016}. 

In actual experiments, quantum gases are confined by trapping potentials that may reduce the low energy dynamics to one dimension under suitable conditions.
The dimensional crossover can be realized
by introducing in a 3D system two counterpropagating laser-beams creating,
by interference, one-dimensional (1D) light wires. The geometrical variation can be driven just
by tuning the laser-beams intensity.
However, two-body collisions conserve a 3D nature at small distance and the connection between 3D and 1D predictions must be analyzed 
with great care. In the quasi-1D regime, namely when essentially one transverse mode is populated, it has been shown for two-component fermions that the 1D 
contact is related to the 3D contact by a simple geometric factor depending on the radial confinement \cite{Valiente2012,Zhou2017}. Another fundamental question 
about the dependence of Tan's contact on dimensionality 
can be formulated this way: What is the behavior of the contact when going from the quasi-1D regime to the 3D regime, that is, when the transverse excited 
states are more and more populated?

In this paper we study the behavior of the 1D Tan's contact in a dilute gas of bosons in the 3D-1D crossover at zero temperature. We provide an analytical 
expression that describes the full crossover from the 3D gas to the quasi-1D regime, and show that it displays clearly distinct behaviors in these two cases. 
As in the quasi-1D case, we show that in the 3D case the 1D and 3D contacts are also related by a geometrical factor.
When transverse fluctuations are negligible, we compare our results with strictly 1D results provided by Lieb-Liniger theory.

The paper is organized as follows. In Section \ref{sec:Model} we present the model used to describe the weakly interacting gas of bosons at zero 
temperature and its conditions of validity. We derive a first formula for Tan's contact in the absence of a longitudinal trap. Then, we use the 
Local Density Approximation (LDA) in order to obtain Tan's contact in the experimentally relevant situation of a cigar-shaped trapped gas in Section 
\ref{sec:Cigar}. Finally, we summarize our results and conclude in Section \ref{sec:Concl}.

\section{Model and quantities of interest}
\label{sec:Model}

The considered system is a weakly interacting Bose gas confined in a cylindrically symmetric 3D harmonic potential at zero temperature. In this regime, 
interactions are accurately described by a contact potential of the form $V(\vec r_1-\vec r_2)=g\delta(\vec r_1-\vec r_2)$ \footnote{Note that this expression for the potential is valid in the Gross-Pitaevskii approach, but the correct formulation for the two-body problem is given by the pseudopotential
  $V(\vec r)\Phi(\vec r)=g\delta(\vec r)\dfrac{\partial}{\partial r}
  ( r\Phi(\vec r))$ \cite{Huang1957,Braaten2006}. }
which is characterized by a 
single parameter, namely the s-wave scattering length $a$, through $g=4\pi\hbar^2 a/m$ \cite{Pitaevskii_book}. At very low densities ($n a^3\ll 1$ with $n$ 
being the three-dimensional atomic density) and for moderate anisotropies of the trapping potential, the ground state of the system is a Bose-Einstein 
condensate (BEC) that obeys the so-called Gross-Pitaevskii equation \cite{Pitaevskii_book}.
In free space, this ground state is simply the homogeneous solution which has for 
momentum distribution $n\delta(\vec k)$. However, when quantum fluctuations are taken into account, Bogoliubov theory shows that the condensate is 
depleted due to interactions and that the tail of the momentum distribution behaves as $\mathcal C/k^4$ \cite{Pitaevskii_book,Clement2016}. As the 
transverse confinement is increased compared to the longitudinal one, the system dynamics is more and more reduced to a single spatial dimension. In the 
extreme situation where only one single transverse mode is populated, the quantum gas is described by the Lieb-Liniger 1D model 
\cite{LiebLiniger} with an interaction parameter $\gamma=-2/(a_{1D}n_1)$ where $n_1$ is the longitudinal atomic density and $a_{1D}$ the effective 
scattering length whose precise expression is given later \cite{Olshanii03}. Again, this model predicts that the tail of the 
momentum distribution behaves as $1/k^4$ for large momenta $\hbar k$. In the following section, we show that the gap between these two regimes 
in a cigar-shaped geometry can be filled by an effective theory. Our approach, which is valid for weak interactions, can be complemented by already known 
results from Lieb-Liniger theory.

\subsection{Effective one-dimensional Gross-Pitaevskii equation for a weakly interacting Bose gas}

In practice, the Bose gas is confined in a 3D harmonic trap of the form 
$U(\vec r)=\frac{1}{2}m\omega_\perp^2 \vec r_\perp^{\,2}+\frac{1}{2}m\omega_z^2 z^2$ where $\vec r_\perp$ and $z$ stand for the radial and longitudinal 
coordinates of an atom. In the limit of a highly anisotropic trap, namely when $\omega_\perp\gg\omega_z$, the transverse confinement is such that an effective 1D dynamics can
be reached. The system is thus accurately described by a 1D order parameter $\psi(z,t)$, depending on a single spatial coordinate $z$ 
along the axial direction of the trap, which obeys the following Gross-Pitaevskii equation \cite{Leboeuf2002,Gerbier2004,Delgado2006}
\begin{equation}
  \label{eq_GP}
i \hbar\frac{ \partial \psi(z,t)}{\partial t}=\left[-\frac{\hbar^2}{2m}\frac{\partial^2}{\partial z^2}+U(z)+\epsilon(n_1)\right]\psi(z,t).   
\end{equation}
Here $n_1(z,t)=|\psi(z,t)|^2$, $U(z)$ is an external potential that will be taken to be either zero or harmonic $U(z)=\frac{1}{2}m\omega_z z^2$ and the 
non-linear term $\epsilon(n)$ describes effective interactions in 1D along the crossover from 3D to 1D. Although its exact analytical expression
is not known, a very accurate approximation is given by~\cite{Fuchs2003,Gerbier2004}
\begin{equation}
\label{mu}
\epsilon(n_1)=\hbar\omega_{\perp}\sqrt{1+4an_1}.
\end{equation}
This equation interpolates between two well known regimes. For low densities ($a n_1\ll 1$), $\epsilon(n_1)\simeq \hbar\omega_\perp+2\hbar\omega_\perp 
a n_1$, the standard one-dimensional Gross-Pitaevskii equation is recovered. In this case the transverse wave function is simply the ground state of the 
harmonic oscillator. In the opposite limit ($a n_1\gg 1$), $\epsilon(n_1)\simeq 2\hbar\omega_\perp \sqrt{a n_1}$ and many transverse modes are occupied. 
These cases are respectively the so-called 1D Mean-Field (MF) and Transverse Thomas-Fermi (TTF) regimes. Later, we will recall some standard solutions of this equation in order to compute the Tan's 
contact in experimentally relevant situations.

\subsection{Bogoliubov approach for Tan's contact}

We now turn to the calculation of Tan's contact from the tails of the momentum distribution. While very accurate for describing the classical dynamics of a 
quasi-1D BEC, Eq.~(\ref{eq_GP}) does not describe correlations between particles which are essential for the existence of an algebraic tail 
in the momentum distribution. Being a mean field equation, it has to be complemented  with quantum fluctuations which we treat at the Bogoliubov level in 
the spirit of \cite{Gerbier2004}. Note however that we do not need to take into account phase fluctuations, although in 1D they destroy long-range order 
even at zero temperature. Indeed, they mainly affect the far off-diagonal part of the one-body-density matrix \cite{Pitaevskii_book}, or, equivalently, the 
low-momentum behavior which is not our interest in this study. 

The Bogoliubov approach \cite{Pitaevskii_book} includes small quantum fluctuations $\delta \hat \psi$ to the condensate wave function $\psi(z)$ (the ground state of the stationary version of (\ref{eq_GP})). A simple way to obtain their equations of motion is to linearize (\ref{eq_GP}) around $\psi$ and use canonical quantization rules for $\delta \hat \psi$ and $\delta \hat \psi^\dagger$. In free space, these fluctuations are expanded on the basis of free particle operators $a_k$ and $a_k^\dagger$ which respectively destroy and create a particle with momentum $\hbar k$. Finally these equations are solved by introducing the quasi-particle operators $ b_k=u_k a_k+v_{-k}a^\dagger_{-k}$, with bosonic commutation relations $[b_k,b_q]=0$ and $[b_k,b_q^\dagger]=\delta_{kq}$. The coherence functions $u_k$ and $v_k$ then obey the following Bogoliubov-de Gennes equations
\begin{equation}
  \label{eq:bog}
  \hbar\omega 
  \begin{pmatrix}
    u_k \\ v_k
  \end{pmatrix}
  =
  \begin{pmatrix}
    \frac{\hbar^2k^2}{2m}+mc^2 & mc^2 \\ -mc^2 &-\frac{\hbar^2k^2}{2m}-mc^2
  \end{pmatrix}
 \begin{pmatrix}
    u_k \\ v_k
  \end{pmatrix}
\end{equation}
where $c$ is the effective sound velocity given by~\cite{Stringari1996,Zaremba1998}
\begin{equation}
  mc^2=n_1\left.\frac{\partial \epsilon}{\partial n}\right|_{n_1}.
\end{equation}
Solving these equations, complemented with $u^2_k-v_k^2=1$ which comes from bosonic commutation relations, yields the so-called Bogoliubov spectrum $\hbar\omega(k)=\sqrt{\hbar^2k^2c^2+(\hbar^2k^2/2m)^2}$ \cite{Bogoliubov1947}.
The distribution of particles with momentum $k\ne 0$ is defined as
\begin{equation}
  n(k)=\langle a_k^\dagger a_k\rangle=(u_k^2+v_k^2)\langle b^\dagger_k b_k\rangle+ v_k^2,
\end{equation} 
with $\langle b^\dagger_k b_k\rangle$ assumed to be Bose-Einstein distributed. At zero temperature, the ground state of the system is the vacuum of 
quasi-particles and  $n(k)=v_k^2$ which corresponds to the quantum depletion of the condensate due to interactions. Solving Eq.~\eqref{eq:bog} 
yields
\begin{equation}
\label{mom}
n(k)=\frac{\hbar^2k^2/2m+mc^2}{2\hbar\omega(k)}-\frac{1}{2},
\end{equation}
Taking the large $k$ limit, one obtains the high-momentum behavior of the momentum distribution
\begin{equation}
\mathcal{C}_{\mathrm{cig}}= \lim_{k\rightarrow\infty}n(k)k^{4},
\end{equation}
where $\mathcal{C}_{\mathrm{cig}}$ is the homogeneous contact density in
the cigar-shaped geometry, given by
\begin{equation}
\label{c1d}
\mathcal{C}_{\mathrm{cig}}=\frac{4}{a_{\perp}^4}\frac{a^2n_1^2}{1+4an_1},
\end{equation}
where $a_{\perp}=\sqrt{\hbar/m\omega_{\perp}}$ is the radial harmonic oscillator length. In the TTF regime, $an_1\gg1$, one gets
\begin{equation}
\label{ctf}
\mathcal{C}_{\mathrm{cig}}^{TTF}=\frac{an_1}{a_{\perp}^4},
\end{equation}
and in the low density regime, $an_1\ll1$, 
\begin{equation}
\label{cMF}
\mathcal{C}_{\mathrm{cig}}^{MF}=\frac{4a^2n_1^2}{a_{\perp}^4}.
\end{equation}
Along the crossover, we observe  a transition from linear to quadratic behavior for the contact density with respect to the 1D atomic density $n_{1}$.

However, one has to be careful to the fact that the mean field approach can no longer describe the system when the quantum correlations are too important, 
i.e when the density is too low (typically, when $a_{\perp}^2n_1/a\precsim 1$ \cite{Stringari2002}). In this case, the physics is described by Lieb-Liniger 
theory \cite{LiebLin}. Nevertheless, there is an intermediate regime described by both theories, allowing us to test our formula (\ref{c1d}).

\subsection{Comparison with Lieb-Liniger theory} 
\label{secLL}
In a strictly 1D situation, the Bose gas is 
accurately described by the Lieb-Liniger equations \cite{LiebLin}.
This model is exactly solvable with the Bethe ansatz and allows us to 
obtain analytical expressions for the contact for both strong and weak values of the interaction parameter. In the following we  recall these 
results and compare them with our prediction in the MF regime for weak interactions.

Tan's adiabatic sweep theorem in 1D allows us to express the 1D contact density $\mathcal{C}_{1D}$ as a function of the adimensional ground state energy via \cite{Olshanii03,Zwerger2011,Lang2017}
\begin{equation}
\mathcal{C}_{1D}=\frac{4n_1^2}{a_{1D}^2}e'(\gamma),
\end{equation}
where $\gamma=-\frac{2}{a_{1D}n_1}$ is the adimensional coupling strength and 
\begin{equation}
\label{a1d}
a_{1D}\simeq \frac{a_{\perp}^2}{a}\left(1-1.03\frac{a}{a_{\perp}}\right)
\end{equation}
is the 1D scattering length \cite{Olshanii1998}. 
Expressions for $e(\gamma)$ in the strong ($\gamma\gg 1$) and weak ($\gamma\ll 1$) coupling 
regimes are provided by Lieb-Liniger (LL) theory \cite{LiebLin}. The weak coupling regime $a_{1D}n_1\gg 1$ can be compared with the MF regime in Eq.~(\ref{cMF}), 
corresponding to $an_1\ll1$. Note that we can consider simultaneously these two limits only if $a_{\perp}\gg a$ (it is sufficient to have $a_{\perp}>10a$ \cite{Stringari2002}). In this case, $a_{1D}\simeq a_{\perp}^2/a$ 
and $e(\gamma)\simeq \gamma$, which yields
\begin{equation}
    \label{cll}
    \mathcal{C}_{1D}^{LL}\underset{a_{1D}n_1\gg 1}{=}\frac{4a^2n_1^2}{a_{\perp}^4}\underset{an_1\ll 1}{=}\mathcal{C}_{\mathrm{cig}}^{MF}.
\end{equation}
Thus, our results for the contact obtained from Bogoliubov and Gross-Pitaevskii theory are compatible with Lieb-Liniger theory.
This is indeed expected since the Lieb-Liniger model is well approximated by the one-dimensional mean field Gross-Pitaevskii 
equation for weak interactions ($(a/a_\perp)^2\ll a n_1 \ll 1$) \cite{LiebLiniger}.

For completeness we now discuss the opposite Strongly Interacting (SI) regime, which is out of the scope of the effective 1D Gross-Pitaevskii
equation. If $a_{1D}n_1\precsim 1$, one enters the strongly correlated regime, where many excited states are populated and one can no longer apply the
Bogoliubov approach. Instead, Lieb-Liniger strong coupling expansion $e(\gamma)\simeq\frac{\pi^2}{3}\left(1-\frac{4}{\gamma}+\frac{12}
{\gamma^2}\right)$ allows us to obtain the following contact density:
\begin{equation}
  \begin{split}
    \label{ctonks}
    \mathcal{C}_{1D}^{LL}& \underset{a_{1D}n_1\ll 1}{=}4\pi^2n_1^4\left(\frac{1}{3}+3a_{1D}n_1\right)\\
    & \underset{a_{\perp}\gg a}{=} 4\pi^2n_1^4\left(\frac{1}{3}+\frac{3a_{\perp}^2n_1}{a}\right).
  \end{split}
\end{equation}
\vspace{0.2cm}

\begin{figure}
\begin{center}
  \includegraphics[width=0.95\linewidth]{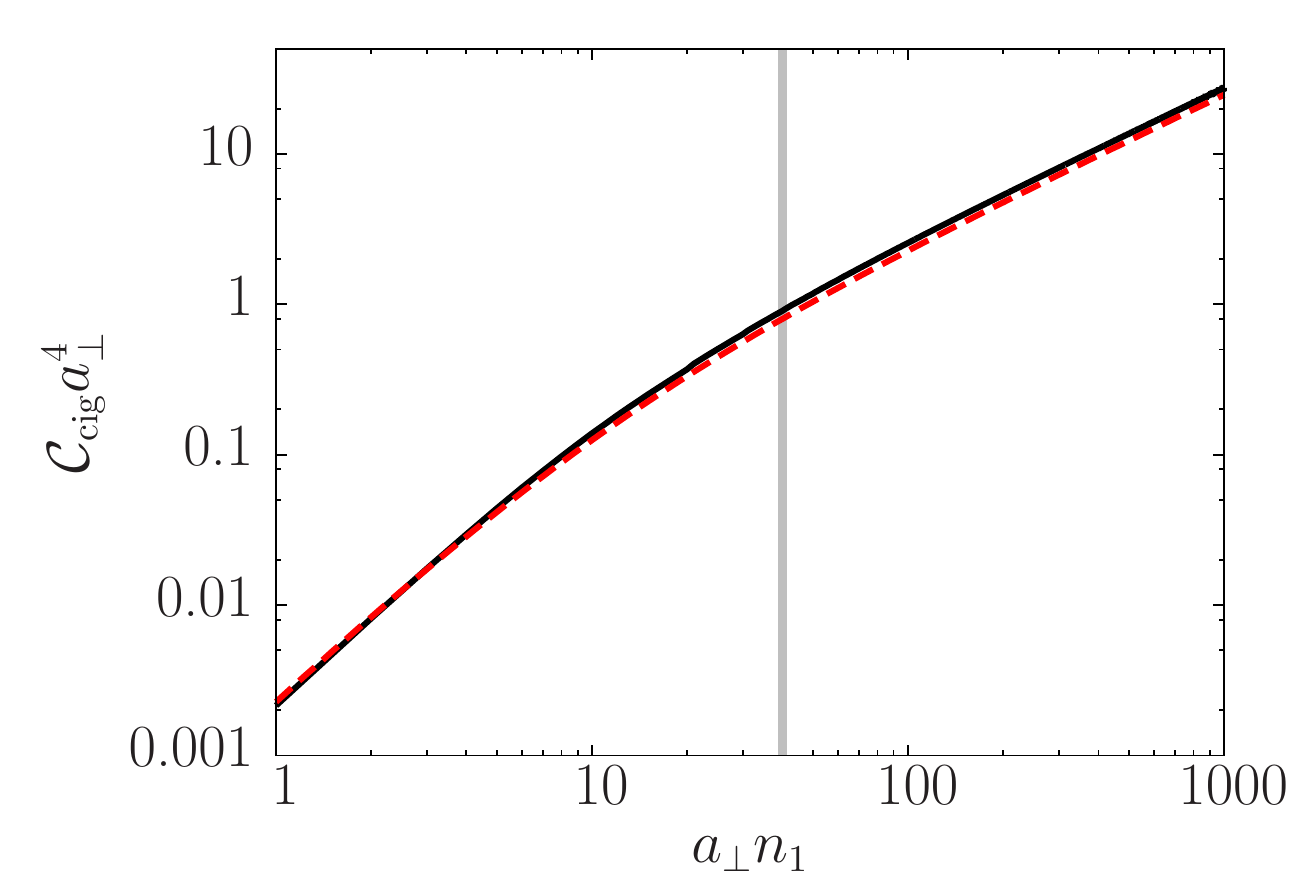}  
  \caption{(Color online.) Contact density $\mathcal{C}_{\mathrm{cig}}$ in units of $a_{\perp}^{-4}$ as a function of $a_{\perp}n_1$, with $a/a_{\perp}=0.025$. The  dashed red line corresponds to our analytic 
  formula Eq.~\eqref{c1d}, the solid black line was numerically obtained from the GLL method. The  vertical gray line represents 
  $an_1=1$ and corresponds to the transition from the MF to the TTF regime.
  \label{fig:gllsound}}
  \end{center}
\end{figure}

In the intermediate interaction regime, the Tan's contact can be obtained by numerically solving the Bethe ansatz equations for the ground state energy $e(\gamma)$. Actually, $e(\gamma)$ beeing computed, it is also possible to take into account the effect of the transverse confinement by mean of a variational ansatz for the $N$-body wavefunction (where $N$ is the number of bosons) of 
the form \cite{Salasnich2004,Salasnich2005}:
\begin{equation}
  \Phi(\vec{r_1},\dots,\vec{r_N})=\Psi(z_1,\dots,z_N)\prod_{i=1}^N\frac{\exp\left(-\frac{x_i^2+y_i^2}{2\sigma^2}\right)}{\sqrt{\pi}\sigma},
  \label{sal1}
\end{equation}
where $\sigma$ is associated with the transverse width of the system in units of $a_{\perp}$. This ansatz leads to the following expression for the total energy per unit length 
\begin{equation}
\label{egll}
\mathcal{E}=\frac{\hbar^2}{2m}n_1^3e\left(\frac{2a}{a_{\perp}^2n_1\sigma^2}\right)+\frac{\hbar\omega_{\perp}}{2}n_1\left(\frac{1}{\sigma^2}+\sigma^2\right),
\end{equation}
which, when minimized with repect to $\sigma$, yields the following implicit equation:
\begin{equation}
\label{sigma}
\sigma^4=1+2an_1e'\left(\frac{2a}{a_{\perp}^2n_1\sigma^2}\right).
\end{equation}
Equations (\ref{sal1})-(\ref{sigma}) describe the Generalized Lieb Liniger (GLL)
approach~\cite{Salasnich2004,Salasnich2005}.
Once Eqs. (\ref{sal1})-(\ref{sigma}) are solved self-consistently,
one can obtain the contact density $\mathcal{C}_{\mathrm{cig}}$ using Eqs.~\eqref{egll} and \eqref{mom} and the formula 
$c=\sqrt{(n_1/m)(\partial^2\mathcal{E}/\partial n_1^2)}$ \cite{Stringari1996,Zaremba1998}. The contacts obtained with this numerical method and with our analytical formula \eqref{c1d}
are plotted as functions of the lineic density $n_1$ in Fig.~\ref{fig:gllsound}. We observe that the curves match, our method having the advantage of leading to a simple analytical formula.

\section{Tan's contact in a cigar-shaped dilute Bose gas}
\label{sec:Cigar}
We now make use of the previous results to describe experimentally relevant situations where the Bose gas is trapped in all directions but with an 
aspect ratio that makes it resembles a cigar. To do so, we make use of the LDA in a harmonic trap of the form 
$U(z)=\frac{1}{2}m\omega_z^2 z^2$ with the proper density profile which depends on the interaction and confinement regimes. We start with a weakly 
interacting Bose gas that is described by the effective 1D equation (\ref{eq_GP}) and then discuss already known results in the strongly 
correlated regime. Then we compare our results with a 3D calculation in the BEC regime.

\subsection{Weakly interacting Bose gas}

In this regime we use the following expression for the atomic density $n_1(z)$ \cite{Delgado2006}
\begin{equation}
\begin{split}
\label{densityprofile}
n_1(z)= & \frac{1}{4a}\left(\frac{\lambda Z}{a_{\perp}}\right)^2\left[ 1-\left(\frac{z}{Z}\right)^2\right] \\
& +\frac{1}{16a}\left(\frac{\lambda Z} {a_{\perp}}\right)^4\left[ 1-\left(\frac{z}{Z}\right)^2\right]^2,
\end{split}
\end{equation}
where $\lambda=\omega_z/\omega_{\perp}$ is the aspect ratio of the trap, and $Z$ is the axial Thomas-Fermi radius obtained from the normalization condition $\int_{-Z}^Zn_1(z)\mathrm{d}z=N$ so that it satisfies the equation:
\begin{equation}
\frac{1}{15}\left(\frac{\lambda Z}{a_{\perp}}\right)^5+\frac{1}{3}\left(\frac{\lambda Z}{a_{\perp}}\right)^3 =\frac{\lambda N a}{a_{\perp}}\equiv\chi_1.
\end{equation}
A good approximation for $Z$ (with a residual error smaller than $0.75\%$) is given by \cite{Delgado2008}
\begin{equation}
\frac{\lambda Z}{a_{\perp}}\simeq\left(\frac{1}{(15\chi_1)^{4/5}+\frac{1}{3}}+\frac{1}{57\chi_1+345}+\frac{1}{(3\chi_1)^{4/3}}\right)^{-1/4}.
\end{equation}
Then, integrating Eq.~(\ref{c1d}) over this profile yields a Tan's contact $\mathcal{I}_{\mathrm{cig}}$ equal to
\begin{widetext}
  \begin{equation}
    \label{ctrap}
    \begin{split}
      \mathcal{I}_{\mathrm{cig}} = & \left\{
        \lambda Z\sqrt{\lambda^2 Z^2 +2a_{\perp}^2}\left(2\lambda^6Z^6+14a_{\perp}^2Z^4\lambda^4+5a_{\perp}^4\lambda^2Z^2-15a_{\perp}^6\right)\right.\\
      & \left. +30a_{\perp}^8
        \operatorname{artanh}\left(\frac{\lambda Z}{\sqrt{\lambda^2Z^2+2a_{\perp}^2}}\right)\right\}\frac{1}{30 a_{\perp}^8
        \lambda\left(\lambda^2 Z^2+2a_{\perp}^2\right)^{3/2}}.
    \end{split}
  \end{equation}
\end{widetext}
Together with Eq.~\eqref{c1d}, this formula constitutes the central result of this paper. Note that the validity of the LDA is guaranteed when $Z\gg a_z$, where $a_z=\sqrt{\hbar/m\omega_z}$ is the transverse harmonic oscillator length.

\begin{figure}
  \begin{center}
    \includegraphics[width=0.95\linewidth]{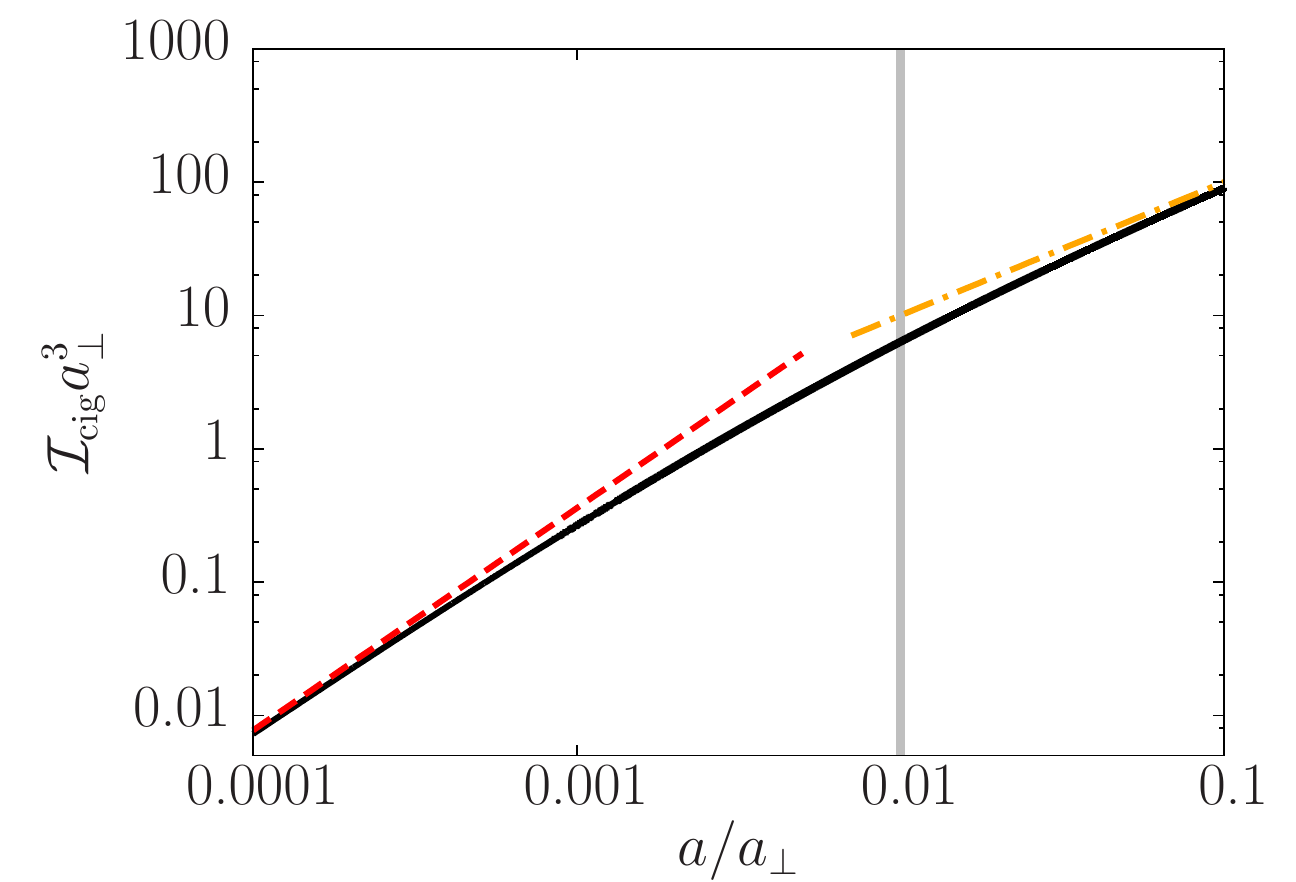}  
    \caption{(Color online.) Tan's contact $\mathcal{I}_{\mathrm{cig}}$ in units of $a_{\perp}^{-3}$ as a function of $a/a_{\perp}$. The black curve 
    represents the total contact from Eq.~(\ref{ctrap}), while the  dashed red and dot-dashed orange curves are $\mathcal{I}_{\mathrm{cig}}^{MF}$ and 
    $\mathcal{I}_{\mathrm{cig}}^{TTF}$ from Eqs (\ref{iMF}) and (\ref{itf}), respectively. The  vertical gray line verifies 
    $\chi_1=N\lambda a/a_{\perp}=1$ and represents the transition from the MF to the TTF regime. The set of parameters used in these plots is 
    $\left(N=1000,\lambda=0.1\right)$.\label{fig:crossover}}
  \end{center}
\end{figure}

As in the previous section, we can consider the limiting cases of the TTF regime and the MF regime. Here, the correct parameter 
governing the transition is $\chi_1$ \cite{Stringari2002}. If $\chi_1\gg1$, one enters the TTF regime, where
\begin{equation}
\label{itf}
\mathcal{I}_{\mathrm{cig}}^{TTF}=\frac{aN}{a_{\perp}^4},
\end{equation}
which is very similar to the homogeneous result (\ref{ctf}). In the opposite MF regime $\chi_1\ll1$ we get
\begin{equation}
\label{iMF}
\mathcal{I}_{\mathrm{cig}}^{MF}=\frac{4(3\lambda)^{\frac{2}{3}}}{5a_{\perp}^{\frac{14}{3}}}a^{\frac{5}{3}}N^{\frac{5}{3}}.
\end{equation}
As in the homogeneous case of Eq.~(\ref{cll}), this formula is in agreement with the result obtained from the LDA performed on weak coupling Lieb-Liniger expansion of the contact density \cite{Olshanii03,Lang2017}. Expressions (\ref{ctrap}), (\ref{itf}) and (\ref{iMF}) for Tan's contact are plotted in Fig. \ref{fig:crossover}.

\begin{figure}
\begin{center}
  \includegraphics[width=0.95\linewidth]{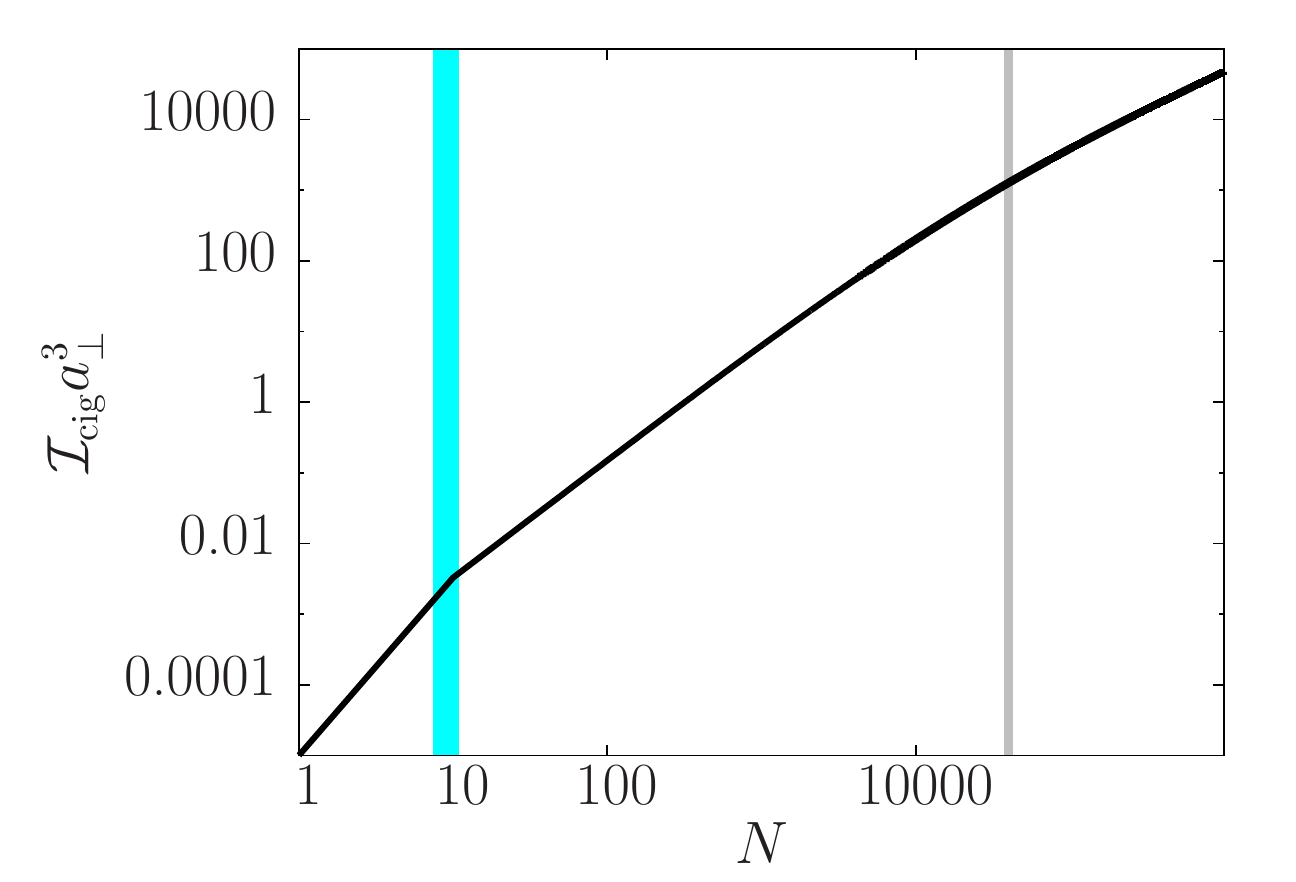}  
  \caption{(Color online.) Tan's contact $\mathcal{I}_{\mathrm{cig}}$ (black line)  in units of $a_{\perp}^{-3}$ as a function of the number of bosons $N$. 
  The thick vertical cyan line represents
  $\xi_1=N \lambda a_{\perp}^2/a^2 = 1$ and corresponds to the transition from the strongly interacting regime to the mean-field regime. The left part is 
  plotted using Eq.~(\ref{itonks}), the right part
  using Eq.~(\ref{ctrap}). Note that there is a small shift at the transition, which is due to corrections to Eq.~(\ref{itonks}) at intermediate values 
  of $a$ (which are not known analytically). The vertical gray
  line corresponds to the transition from the MF regime to the TTF regime at $\chi_1=N\lambda a/a_{\perp}=1$.
   The set of parameters used in these plots is
  $\left(\lambda=0.0005,a/a_{\perp}=0.05\right)$.\label{fig:contactn}}
  \end{center}
\end{figure}

\subsection{Strongly interacting gas}

% \begin{widetext}
\begin{table*}
\renewcommand{\arraystretch}{2}
\begin{center}
\begin{tabular}{|c||c|c|c|c|c|}
\hline 
Regime & TTF & TTF $\to$ MF & MF & MF $\to$ SI & SI \\\hline\hline
Parameters & $\chi_1\gg 1$ & $\chi_1\simeq 1$ &  $\chi_1\ll 1$ and $\xi_1\gg 1$ & $\xi_1\simeq 1$ & $\xi_1\ll 1$\\\hline
$n_1(z)$ & $\propto (1-z^2/Z^2)^2$ & Eq.~(\ref{densityprofile}) & $\propto (1-z^2/Z^2)$ & * & $\propto (1-z^2/Z^2)^{1/2}$\\\hline
$\mathcal{C}$ & $\frac{an_1}{a_{\perp}^4}$ & Eq.~(\ref{c1d}) & $\frac{4a^2n_1^2}{a_{\perp}^4}$ & *& $4\pi^2n_1^4\left(\frac{1}{3}+3a_{1D}n_1\right)$ \\\hline
$\mathcal{I}$ & $\frac{aN}{a_{\perp}^4}$ & Eq.~(\ref{ctrap}) & 
$\frac{4(3\lambda)^{\frac{2}{3}}}{5}a_{\perp}^{-\frac{14}{3}}a^{\frac{5}{3}}N^{\frac{5}{3}}$ & *
& Eq.~(\ref{itonks})\\\hline
\end{tabular}\caption{Summary of the results for the contact density $\mathcal{C}$ and the integrated contact $\mathcal{I}$ in the different regimes, with the 
associated density profiles and the parameters governing the transition in the cigar-shaped dilute Bose gas. The transition parameters are given by $\chi_1=N\lambda a/a_{\perp}$ 
and $\xi_1=N\lambda a_{\perp}^2/a^2$. The asterisk (*) indicates that there is no analytical formula in the literature, and that one should solve numerically the Bethe Ansatz
equations of the Lieb-Liniger model.}\label{summary}
\end{center}
\end{table*}
% \end{widetext}

Again, for the sake of completeness we now discuss the SI regime in the strictly one-dimensional situation. The relevant parameter that 
describes the transition between the mean-field and the strongly correlated regimes is now given by
$\xi_1\equiv Na_{1D}^2/a_z^2=N\lambda a_{\perp}^2/a^2$ \cite{Stringari2002,PetrShlyap00,Olshanii2001}. In the SI regime $\xi_1\ll 1$, 
 the density profile is well described by the Tonks (fermionic) one given by $n_1^{\mathrm{Tonks}}(z)=\frac{\sqrt{2N\lambda}}{\pi a_{\perp}}
\sqrt{1-\frac{z^2}{{Z_{\mathrm{Tonks}}}^2}}$ with $Z_{\mathrm{Tonks}}=\sqrt{\frac{2N}{\lambda}}a_{\perp}$. In order to obtain the Tan's contact, 
finite $g$ corrections are included in the density profiles by consistently solving Lieb-Liniger equations \cite{Decamp2016b,Lang2017}. This yields
\begin{equation}
\label{itonks}
\mathcal{I}_{1D}^{\mathrm{SI}}=\frac{N^{\frac{5}{2}}\lambda^{\frac{3}{2}}}{a_{\perp}^3}\left[\frac{256\sqrt{2}}{45\pi^2}+\frac{a_{1D}
\sqrt{N\lambda}}{a_{\perp}}\left(\frac{70}{9\pi^2}-\frac{8192}{81\pi^4}\right)\right].
\end{equation}

The transition from the SI regime to the MF and TTF regimes is plotted for the contact as a function of the number of bosons in Fig. 
\ref{fig:contactn}. At a low number of particles, 
the system is in the SI regime and the contact is given by Eq.~\eqref{itonks}. When $N$ is increased the system enters the 
MF regime (thick vertical green line) when $\xi_1=N \lambda a_{\perp}^2/a^2 = 1$ and the contact is accurately described by Eq.~(\ref{iMF}). 
Upon further increasing $N$, when $\chi_1=N\lambda a/a_{\perp}\ge1$ (vertical cyan line), the system is essentially 3D and the contact is given by Eq.~\eqref{itf}.

The different regimes and associated contacts are summarized in Table \ref{summary}.

\subsection{Comparison with the 3D contact}

It is important to note that, until now, we have considered \textit{one-dimensional contacts} (homogeneous to [length$^{-3}$]), in order to compare our results 
with strictly 1D results provided by Lieb-Liniger theory. Since ultracold experiments are done in a 3D world, it is natural to ask how these quantities are related 
to the \textit{three-dimensional contacts} (homogeneous to [length$^{-1}$]). In the quasi-1D case, which corresponds in our system to the MF and SI regimes, 
it has been shown that it
is sufficient to multiply the 1D
contact by a geometric factor of the type $\pi d^2$, where $d$ is the transverse radius of the cigar (and is equal to $a_{\perp}$ in 
the quasi-1D regime) \cite{Valiente2012,Zhou2017}: 
\begin{equation}
\label{3d1d}
\mathcal{I}_{3D}=\pi d^2\mathcal{I}_{1D}.
\end{equation}
Interestingly, although the system is highly non-homogeneous in the transverse direction, Eq.~\eqref{3d1d} shows that everything behaves as if it were a cylinder of radius 
$d$ with a constant 1D-contact (or 
contact lineic density)  in the radial direction.

In the TTF regime however, as there are many transverse states, we expect that the transverse non-uniformities 
will no longer be negligible when comparing the 3D and 1D contacts. In this paragraph we compare Eq.~(\ref{itf}) with the 3D contact in a 3D dilute Bose gas obtained in \cite{Clement2016}. In this case the 
Bogoliubov approach yields
\begin{equation}
\label{chom}
\mathcal{C}_{3D}=\frac{m^4c_{3D}^4}{\hbar^4},
\end{equation}
where $c_{3D}$ is the 3D velocity of sound. Supposing that the Thomas-Fermi regime is reached in the three
directions, we integrate Eq.~(\ref{chom}) in the LDA  using 
\begin{equation}
c_{3D}^2 =\frac{4\pi\hbar^2an_0}{m^2}\left(1-\frac{x^2}{R_x^2}-\frac{y^2}{R_y^2}-\frac{z^2}{R_z^2}\right),
\end{equation}
where $R_i$ is the Thomas-Fermi radius along direction $i$ and $n_0$ is the density at the center of the trap. This leads to
\begin{equation}
\mathcal{I}_{3D}=\frac{64\pi^2}{7}a^2Nn_0.
\end{equation}
Using the expression of $n_0$ in a cigar-shaped dilute Bose gas \cite{Baym1996} one gets
\begin{equation}
\label{i3d}
\mathcal{I}_{3D}=\frac{8\pi}{7}\frac{aN}{a_{\perp}^2}\left(\frac{15\lambda N a}{a_{\perp}}\right)^{\frac{2}{5}}.
\end{equation}
 If we consider
$R_{\perp}=R_x=R_y=2a_{\perp}(an_1(0))^{1/4}$ \cite{Pitaevskii_book}, we find
\begin{equation}
\mathcal{I}_{3D}=\frac{8}{7}\pi R_{\perp}^2\frac{aN}{a_{\perp}^4}.
\end{equation}
The quantity $S=\frac{8}{7}\pi R_{\perp}^2$ is a purely geometrical factor equal to the cross section in the center of the trap up to a numerical factor $8/7$
which is due to the high non-uniformity of the system. Thus, we have found:
\begin{equation}
\mathcal{I}_{3D}=S\mathcal{I}_{\mathrm{cig}}^{TTF}.
\end{equation}
This formula has the same form as Eq.~\eqref{3d1d} and constitutes its generalization in the TTF regime.

\section{Conclusions}
\label{sec:Concl}

In this work we have studied the weight of the tail of the momentum distribution, Tan's contact, for a zero 
temperature dilute Bose gas trapped in a cigar-shaped potential.
We have derived an analytical expression that describes the whole crossover from the 3D Thomas-Fermi regime to the 1D mean-field regime, which corresponds 
to a progressively decreasing  number of transverse excited states. In the latter case, our formula is in perfect agreement with strictly 1D results derived 
using Lieb-Liniger theory. In the 3D regime, we have compared this 1D contact with the 3D contact obtained by LDA in the Bogoliubov approach, and have shown that these are related by a geometric factor.

Our expression for Tan's contact is expressed in terms of controllable experimental parameters such as the aspect ratio of the trap, the scattering length and the number 
of trapped bosons. It displays very distinct dependences on these parameters in the 3D and 1D regimes. In this sense, Tan's contact could be a useful 
experimental observable characterizing the dimensional regime of the system. Moreover, this work provides corrections to strictly 1D results, and, 
thus, can be a guide for actual experiments devoted to Tan's contact measurements.

\begin{acknowledgments}
We aknowledge L. Salasnich for useful discussions.
\end{acknowledgments}

%%%%%%%%%%%%%%%%%%%%%%%%%%%%%%%%%%%%%%%%%%%%%%%%%%%%%%%%%%%%%%%%%%%%%%%%%%%%
%\bibliographystyle{prsty}
%\bibliography{biblioferm}

\begin{thebibliography}{10}

\bibitem{Tan2008a}
S. Tan, Ann. Phys. (N.Y.) {\bf 323},  2971  (2008).

\bibitem{Tan2008b}
S. Tan, Ann. Phys. (N.Y.) {\bf 323},  2987  (2008).

\bibitem{Tan2008c}
S. Tan, Ann. Phys. (N.Y.) {\bf 323},  2952  (2008).

\bibitem{Decamp2016b}
J. Decamp, J. Junemann, M. Albert, M. Rizzi, A. Minguzzi, and P. Vignolo,
  Physical Review A {\bf 94},  053614  (2016).

\bibitem{Decamp2017}
J. Decamp, J. J\"unemann, M. Albert, M. Rizzi, A. Minguzzi, and P. Vignolo, New
  Journal of Physics {\bf 19},  125001  (2017).

\bibitem{Clement2016}
R. Chang, Q. Bouton, H. Cayla, C. Qu, A. Aspect, C.~I. Westbrook, and D.
  Cl\'ement, Phys. Rev. Lett. {\bf 117},  235303  (2016).

\bibitem{Valiente2012}
M. Valiente, N.~T. Zinner, and K. M\o{}lmer, Phys. Rev. A {\bf 86},  043616
  (2012).

\bibitem{Zhou2017}
M. {He} and Q. {Zhou}, arXiv:1708.00135.

\bibitem{Huang1957}
K. Huang and C.~N. Yang, Phys. Rev. {\bf 105},  767  (1957).

\bibitem{Braaten2006}
E. Braaten and H.-W. Hammer, Phys. Rep. {\bf 428},  259  (2006).

\bibitem{Pitaevskii_book}
L. Pitaevskii and S. Stringari, {\em Bose-Einstein Condensation and
  Superfluidity} (Oxford University Press, Oxford, UK, 2016).

\bibitem{LiebLiniger}
E. Lieb, Phys. Rev. {\bf 130},  1616  (1963).

\bibitem{Olshanii03}
M. Olshanii and V. Dunjko, Phys. Rev. Lett. {\bf 91},  090401  (2003).

\bibitem{Leboeuf2002}
P. Leboeuf and N. Pavloff, Phys. Rev. A {\bf 64},  033602  (2001).

\bibitem{Gerbier2004}
F. Gerbier, Europhys. Lett. {\bf 66},  771  (2004).

\bibitem{Delgado2006}
A. Mu\~noz Mateo and V. Delgado, Phys. Rev. A {\bf 74},  065602  (2006).

\bibitem{Fuchs2003}
J.~N. Fuchs, X. Leyronas, and R. Combescot, Phys. Rev. A {\bf 68},  043610
  (2003).

\bibitem{Stringari1996}
S. Stringari, Phys. Rev. Lett. {\bf 77},  2360  (1996).

\bibitem{Zaremba1998}
E. Zaremba, Phys. Rev. A {\bf 57},  518  (1998).

\bibitem{Bogoliubov1947}
N. Bogoliubov, J. Phys. USSR {\bf 11},  23  (1947).

\bibitem{Stringari2002}
C. Menotti and S. Stringari, Phys. Rev. A {\bf 66},  043610  (2002).

\bibitem{LiebLin}
E. Lieb and W. Liniger, Phys. Rev. {\bf 130},  1605  (1963).

\bibitem{Zwerger2011}
M. Barth and W. Zwerger, Annals of Physics {\bf 326},  2544  (2011).

\bibitem{Lang2017}
G. Lang, P. Vignolo, and A. Minguzzi, The European Physical Journal Special
  Topics {\bf 226},  1583  (2017).

\bibitem{Olshanii1998}
M. Olshanii, Phys. Rev. Lett. {\bf 81},  938  (1998).

\bibitem{Salasnich2004}
L. Salasnich, A. Parola, and L. Reatto, Phys. Rev. A {\bf 70},  013606  (2004).

\bibitem{Salasnich2005}
L. Salasnich, A. Parola, and L. Reatto, Phys. Rev. A {\bf 72},  025602  (2005).

\bibitem{Delgado2008}
A. Mu\~noz Mateo and V. Delgado, Phys. Rev. A {\bf 77},  013617  (2008).

\bibitem{PetrShlyap00}
D. Petrov, G. Shlyapnikov, and J. Walraven, Phys. Rev. Lett. {\bf 85},  3745
  (2000).

\bibitem{Olshanii2001}
V. Dunjko, V. Lorent, and M. Olshanii, Phys. Rev. Lett. {\bf 86},  5413
  (2001).

\bibitem{Baym1996}
G. Baym and C.~J. Pethick, Phys. Rev. Lett. {\bf 76},  6  (1996).

\end{thebibliography}

\end{document}